\documentclass[authoryear,preprint,review,12pt]{elsarticle}

\usepackage{amssymb}
\usepackage{amsthm}

\usepackage{lineno}

\usepackage{amsmath}
\usepackage{bm}
\usepackage{framed}
\usepackage{xcolor}
\usepackage[noend]{algpseudocode}
\usepackage{algorithmicx,algorithm}
\usepackage{comment}

\journal{arXiv}

\begin{document}

\begin{frontmatter}

\title{CR-Lasso: Robust cellwise regularized sparse regression}

\author[1]{Peng Su}

\author[1]{Garth Tarr}
\author[1,2]{Samuel Muller\corref{cor1}}
\ead{samuel.muller@mq.edu.au}
\author[3]{Suojin Wang}

\cortext[cor1]{Corresponding author}

\affiliation[1]{organization={School of Mathematics and Statistics},
 addressline={The University of Sydney}, 
 state={NSW},
 postcode={2006},
 country={Australia}}

\affiliation[2]{organization={School of Mathematical and Physical Sciences},
 addressline={Macquarie University}, 
 state={NSW},
 postcode={2109},
 country={Australia}}

\affiliation[3]{organization={Department of Statistics},
 addressline={Texas A\&M University}, 
 city={College Station},
 state={Texas},
 postcode={77843}, 
 country={USA}}

\begin{abstract}
Cellwise contamination remains a challenging problem for data scientists, particularly in research fields that require the selection of sparse features. Traditional robust methods may not be feasible nor efficient in dealing with such contaminated datasets. A robust Lasso-type cellwise regularization procedure is proposed which is coined CR-Lasso, that performs feature selection in the presence of cellwise outliers by minimising a regression loss and cell deviation measure simultaneously. The evaluation of this approach involves simulation studies that compare its selection and prediction performance with several sparse regression methods. The results demonstrate that CR-Lasso is competitive within the considered settings. The effectiveness of the proposed method is further illustrated through an analysis of a bone mineral density dataset.

\end{abstract}



\begin{keyword}
Cellwise contamination \sep Cellwise regularization \sep Robust sparse regression \sep feature selection


\end{keyword}

\end{frontmatter}


\section{Introduction}
\label{intro}

Identifying the most important features in an $n\times p$ design matrix $\bm{X}$ to predict an outcome vector $\bm{y}$ is a fundamental problem in statistics. 
This problem becomes challenging when there is contamination in the data. That is, when some elements of the full data matrix $[\bm{y},\bm{X}]$ are corrupted. 
{It is commonly believed that most raw real-data, prior to data cleaning, contain a small proportion of outliers \citep{hampel1986robust}. }
Potential issues caused by these outliers are often ignored \citep{rousseeuw2005robust} even though they may negatively impact estimation and variable selection \citep{weisberg2005applied}.

\begin{figure}[!ht]
	\centering
	\includegraphics[width=12cm]{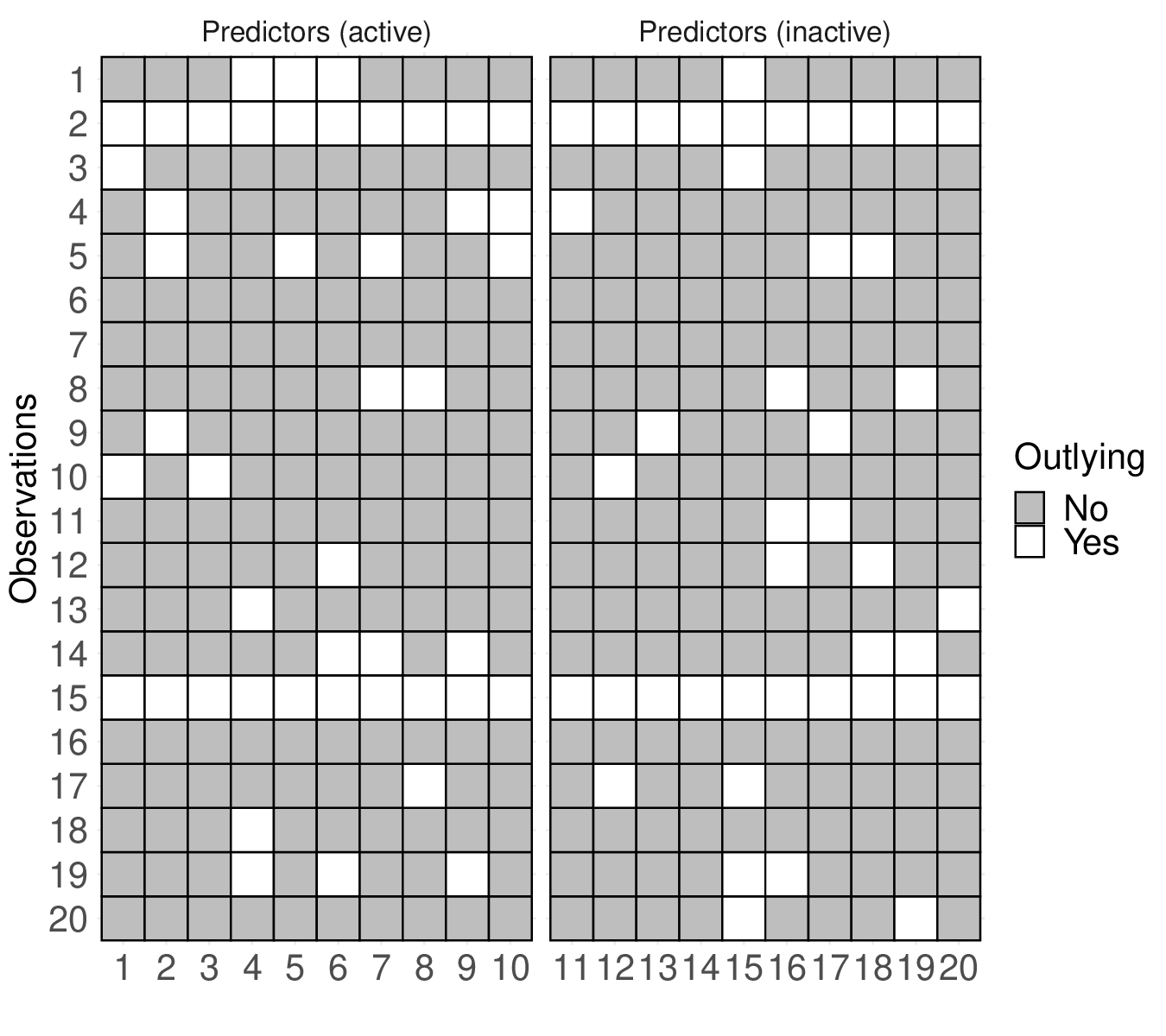}
	\caption{Illustration of outlying cells in the design matrix of a regression model. Rowwise outliers refer to observations whose entire observation vector is contaminated, such as in row 2 and in row 15. Cellwise outliers refer to cells that are contaminated.}
 \label{illustration}
\end{figure}

{There are two typical categories of outliers:} rowwise outliers and cellwise outliers. Rowwise outliers refer to observation vectors whose components are entirely contaminated, such as visualised in row 2 and row 15 in Figure \ref{illustration}. 
One way to overcome this challenge is through robust sparse estimators that can deal with rowwise outliers under sparse settings by combining traditional robust estimators with Lasso-type regularization. For example,
{\citet{alfons_sparse_2013} proposed to incorporate least trimmed squares with the Lasso \citep{Tibshirani_1996}, }
\citet{Smucler_Yohai_2017} and \citet{chang_robust_2018} proposed combining the adaptive Lasso \citep{zou_adaptive_2006} and MM-estimation \citep{Yohai_1987}, respectively. These estimators work by downweighting outlying observations. 

Cellwise outliers refer to cells that are contaminated in a way that makes them different to the true data generating process. It is typically assumed that these outliers occur independently across the design matrix $\bm{X}$. Observations may experience some form of corruption in one or more of the predictors, such as visualized by the scattered white cells in Figure \ref{illustration}.
Cellwise outliers are more challenging than rowwise outliers because of the propagation of these outliers to many of the rows, where even a small proportion of cellwise outliers in each predictor could cause a large proportion of outlying observation vectors \citep{alqallaf_propagation_2009}. 

Outlier detection is often the first step taken when dealing with cellwise outliers in a dataset. One common approach, as suggested by \citet{rousseeuw_detecting_2018}, is to predict the value of a cell and flag it as an outlier if the difference between its predicted value and observed value exceeds a certain threshold. Another technique, described by \citet{debruyne_outlyingness_2019}, treats all outliers as rowwise outliers and identifies the cells that contribute most to the outlying rows. \citet{raymaekers_handling_2021} presented ``cellflagger", a method that detects outlying cells for each row by combining Lasso regularization with a stepwise application of cutoff values.

In the field of robust regression under cellwise contamination, several methods have been proposed to address the issue of outliers. 
For example, \citet{ollerer_shooting_2016} introduced the Shooting S-estimator, which combines the Shooting algorithm and the S-estimator to perform robust regression.
\citet{leung_robust_2016} proposed a three-step regression method that involves detecting rowwise and cellwise outliers, followed by processing covariance matrix estimation and regression.
Alternatively, \citet{filzmoser_cellwise_2020} proposed a cellwise robust M-estimator that replaces the detected outliers with imputed values.

Recently, there has been substantial interest in robust sparse regression under cellwise contamination. 
\citet{Machkour2020} investigated giving adaptive weights for predictors and observations based on detected outliers.
\citet{onur2021robust} proposed to filter and impute outliers before regression modelling to prevent the possible damage caused by outliers. 
\citet{Bottmer_Croux_Wilms_2022} considered solving this problem by running successive penalized S-regression for all predictors and called this method Sparse shooting S (SSS). 

In this paper, we provide a new perspective on dealing with cellwise outliers in regression models.
An active predictor with a cellwise outlier generally increases a regression residual's magnitude and the cell deviation. Building on this idea and the work in \citet{chen2013robust}, we propose the cellwise regularized Lasso (CR-Lasso), which incorporates a regression and cell deviation measure into a loss function. We then  apply a cellwise regularized and sparse regression procedure that helps identify active predictors and possible outliers.

The structure of this paper is as follows. Section ~\ref{method} describes the proposed method and algorithm details. Section~\ref{simu} illustrates the simulation results in low and high-dimensional settings. A real data application is presented in Section~\ref{data}. Finally, we state some conclusions in Section \ref{discussion}. {R functions and sample codes that implement the proposed approach are available on the GitHub page of the first author (https://github.com/PengSU517/regcell).}

\section{Cellwise robust sparse regression}
\label{method}
We first explore the limitations of traditional regularization techniques when dealing with cellwise contamination. We then introduce a novel approach that addresses these issues by providing a constrained loss function, which will be discussed in detail below. 

Consider an observed response $y_i$, a set of $p$ predictors $\bm{x}_i$ and a corresponding coefficient vector $\bm{\beta}$ in a linear regression model framework:
\begin{equation}
 { {y}_{i}={\bm {x}}_{i}^{\top }{\bm {\beta }}+{\varepsilon }_{i}},\ { i=1,\,2,\ldots ,\,n},
 \label{modeleq}
\end{equation}
where in classical settings the error term, $\varepsilon_i$, is assumed to be independent $N(0,\sigma^2)$ distributed. We can write this as $ {\bm {y}}=\bm {X} {\bm {\beta }}+{\bm {\varepsilon }}$, where $\bm y$ is an $n$ dimensional response vector and ${\bm {X}}$ is an $n\times p$ design matrix both of which may include some outliers, {and ${\bm {\varepsilon }}$ is the error vector.} Without loss of generality, unless otherwise specified, we assume all predictors have zero mean and unit variance and the variance of the error term $\varepsilon$ equals one. If not otherwise mentioned, we do not include an intercept term in the linear regression model as we work with centered data. 

\subsection{Modification of the regression loss}

The iterative procedure for outlier detection (IPOD) method \citep{She_Owen_2011} is a modification of Lasso and handles outlying rows by adding an extra term $\bm \zeta$:
\begin{equation}
 \mathop{\operatorname{argmin}}\limits_{\bm\beta, \bm \zeta} \frac{1}{2}\|{\bm y} - {\bm X \bm \beta} - \bm \zeta\|_2^2 + \theta \|\bm \zeta\|_1,
 \label{shenowen}
\end{equation}
where $\bm \zeta$ indicates possible outlying parts in the response $\bm y$ and $\theta$ is a tuning parameter for $\bm \zeta$.
 \citet{She_Owen_2011} showed the equivalence between IPOD and the Huber's M-estimate $\bm{\hat \beta}_H = \mathop{\operatorname{argmin}}\limits_{\bm\beta} \rho_\theta ({\bm y} - {\bm X \bm \beta})$ \citep{huber_robust_2004}, where Huber's loss function is
\begin{equation}
 \rho_\theta(z) = 
 \left\{\begin{array}{ll}
 \theta|z|-\theta^{2} / 2, & \text { if }|z|>\theta,\\ 
 z^{2} / 2, & \text { if }|z| \leq \theta.
 \end{array}\right. 
\end{equation}
With some non-convex penalties, IPOD is equivalent to other M-estimates. {The non-convex penalties in this scenario can be likened to redescending loss functions, known for their robustness against rowwise outliers, including leverage points.} However, like M-estimates, IPOD is only robust against rowwise outliers.

To be robust against cellwise outliers, \citet{chen2013robust} suggested the modification
\begin{equation}
 \mathop{\operatorname{argmin}}\limits_{\bm\beta, \bm \Delta} \frac{1}{2}\|{\bm y} - {(\bm X - \bm \Delta) \bm \beta}\|_2^2 + \eta \|\bm \Delta\|_1,
 \label{cellloss}
\end{equation}
where $\bm \Delta$ is an $n\times p$ matrix, indicating possible outlying parts in the design matrix $\bm X$, and $\eta$ is a tuning parameter for $\bm \Delta$.
However, the solution of (\ref{cellloss}) is non-convex and non-tractable because of the bi-linear term $\bm \Delta \bm \beta$ \citep{chen2013robust}. 

{
In earlier work, \citet{zhu2011sparsity} considered using the squared Frobenius norm $\|\bm \Delta \|_F^2$ as a penalty: 
\[
 \mathop{\operatorname{argmin}}\limits_{\bm\beta, \bm \Delta} \frac{1}{2}\|{\bm y} - {(\bm X - \bm \Delta) \bm \beta}\|_2^2 + \eta \|\bm \Delta\|_F^2,
\]
which is more suitable to deal with measurement errors (where $\bm\Delta$ is dense and bounded) instead of cellwise outliers (where $\bm\Delta$ is assumed to be sparse). 


Addressing cellwise outliers involves individually regularizing cells within the design matrix $\bm{X}$. This leads to the formulation of the following loss function:
\begin{equation}
    \mathop{\operatorname{argmin}}\limits_{\bm \Delta} \frac{1}{2} \|\bm X- \bm \Delta\|_F^2 + \eta \|\bm \Delta\|_1,
    \label{winsor}
\end{equation}
which is solved by
\[
     \hat{\Delta}_{ij} = 
     \left\{
     \begin{array}{ll}
        \text{sign}(x_{ij})(|x_{ij}| - \eta), & \text { if }|x_{ij}|>\eta,\\ 
        0, & \text { if }|x_{ij}| \leq \eta.
    \end{array}\right. 
\]
Then a regularized design matrix $\tilde{\bm { X}} = \bm {X} -\bm {\hat \Delta}$ is obtained by
\[
     \tilde{x}_{ij} = 
     \left\{
     \begin{array}{ll}
        \text{sign}(x_{ij})\eta, & \text { if }|x_{ij}|>\eta,\\ 
        x_{ij}, & \text { if }|x_{ij}| \leq \eta.
    \end{array}\right. 
\]
This solution is equivalent to Winsorization directly \citep{dixon1960simplified}. The similarity between Winsorization and Huber's loss was noted in \citet{Huber1964robust}. Although this connection has not been extensively investigated, a link has previously been established between Huberization and the $L_1$ penalty \citep{sardy2001robust}.

However, using the formula \eqref{winsor}, we can only marginally shrink cells in the design matrix.
To deal with cellwise outliers in a regression model, we propose to combine loss functions \eqref{cellloss} and \eqref{winsor}. The underlying idea is that, in an ideal scenario, a cellwise outlier within an active predictor should have two noticeable effects. Firstly, the regression residual's magnitude should increase, indicating a deviation from the expected relationship. Secondly, it should also contribute to an increased deviation of this cell, highlighting its distinctiveness compared to the other cells in the design matrix.
}

From this view, we modify the loss function as the sum of the regression loss and the deviation loss:
\begin{equation}
 \mathop{\operatorname{argmin}}\limits_{\bm\beta, \bm \Delta} \frac{1}{2}\| \bm y- ( \bm { X} - \bm \Delta) \bm \beta \|_2^2 + \frac{1}{2} \|\bm X- \bm \Delta\|_F^2 + \eta \|\bm \Delta \|_1,
 \label{cellloss2}
\end{equation}
where $\| \bm y- ( \bm { X} - \bm \Delta) \bm \beta \|_2^2$ measures the regression loss and $\|\bm X- \bm \Delta\|_F^2$ measures the deviation loss. The goal of the minimisation problem in \eqref{cellloss2} is to decrease the sum of the regression loss and the deviation loss by shrinking only a few outlying cells in the design matrix. 

However, shrinking cells in predictors can only deal with outliers in $\bm{X}$. Thus, we add a $\bm{\zeta}$ term that addresses possible outliers in $\bm{y}$:
\begin{equation}
 \mathop{\operatorname{argmin}}\limits_{\bm\beta, \bm \Delta, \bm \zeta} \frac{1}{2} \| \bm y- ( \bm { X} - \bm \Delta) \bm \beta - \bm \zeta \|_2^2 + \frac{1}{2} \|\bm X- \bm \Delta\|_F^2 + \eta\|\bm \Delta\|_1+ \theta\|\bm \zeta\|_1,
 \label{cellloss3}
\end{equation}
where $\bm \zeta$ indicates possible outlying parts in the response $\bm y$ and $\theta$ is a tuning parameter to control the sparsity in $\bm \zeta$.

\begin{figure}[!ht]
	\centering
	\includegraphics[width=12cm]{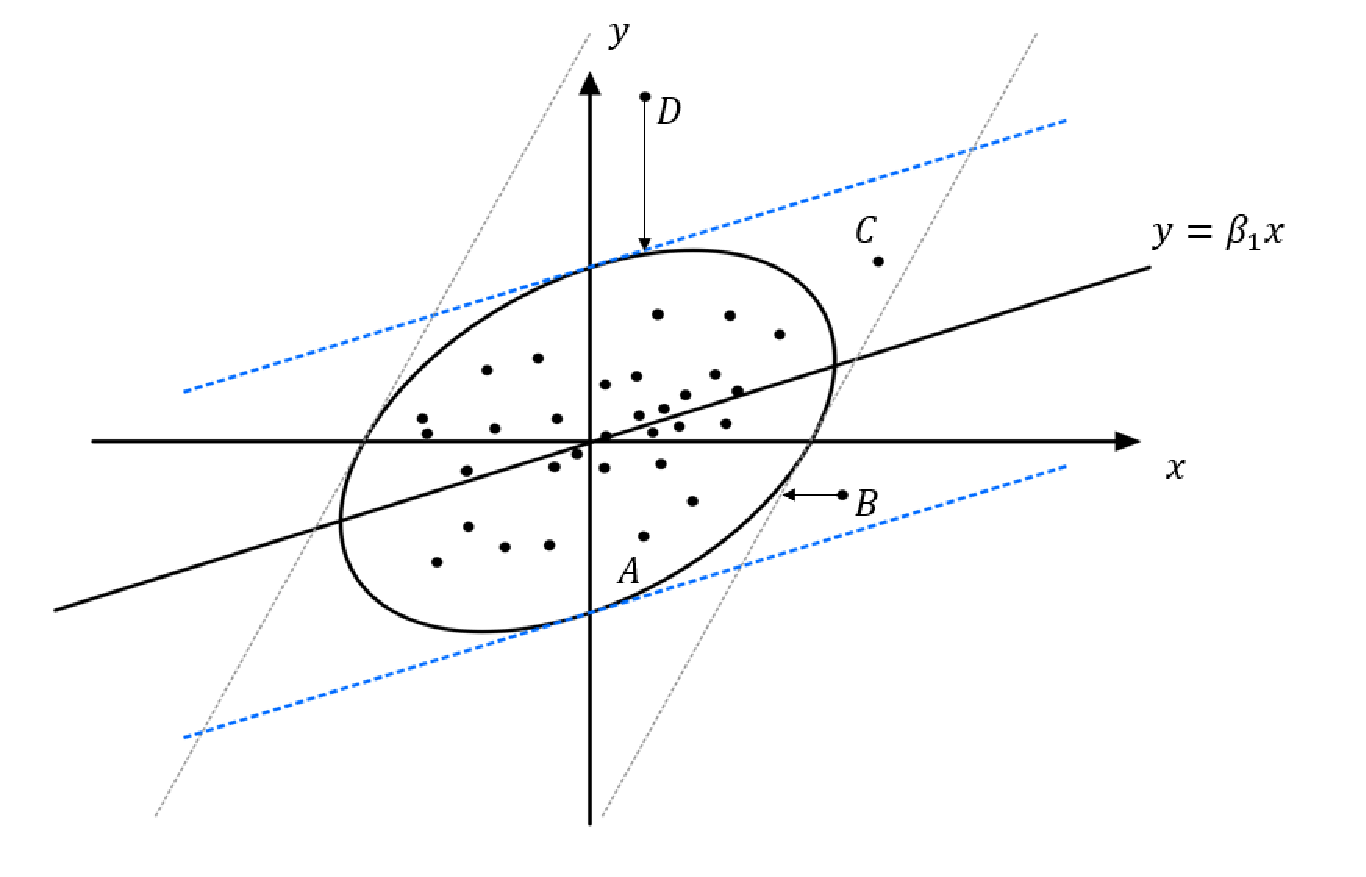}
	\caption{Illustration of cellwise regularization for a simple linear regression model. Grey dashes indicate the boundaries of cellwise regularization (horizontally) in $x$. Blue dashes indicate the boundaries of cellwise regularization (vertically) in $y$.}
	\label{illus_points}
\end{figure}

To illustrate the mechanism, Figure \ref{illus_points} shows a simple linear regression model through the origin: 
$$y_i = \beta_1 x_i + \varepsilon_i, i = 1,\ldots, n.$$

The majority of the clean observations are distributed in the ellipse, such as point A. The grey dashes represent the boundaries of cell regularization in $x$. Any cells outside the grey boundaries will be regarded as outliers in $x$ and {will undergo $\eta$-regularization. This regularization process involves horizontally shifting the outliers towards the grey boundary until they reach it}. Regularizing point B will simultaneously lead to a decrease in both the regression residual and deviation magnitude. Conversely, regularizing point C will reduce the deviation loss but increase the regression loss. In this case, point C will be regarded as a good high-leverage point and will not be regularized, despite having a large deviation magnitude. Blue dashes indicate the boundaries of cellwise regularization in the response $y$. Point D has a large regression residual but a small deviation. In this case, point D will be regarded as an outlier in $y$ and {will be $\theta$-regularized, which means {it} will be moved vertically until they reach the blue boundary.}

{
We observe that the objective loss function \eqref{cellloss3} is convex for $\bm{\zeta}, \bm{\Delta}$, and $\bm{\beta}$, respectively. However, because of the presence of the bilinear term $\bm{\Delta}\bm{\beta}$, it is not jointly convex to all the parameters we considered. Fortunately, to solve the objective loss function \eqref{cellloss3}, an iterative block coordinate descent algorithm can be used, yielding successive estimates of $\bm{\zeta}, \bm{\Delta}$, and $\bm{\beta}$. 

The algorithm can be divided into two stages: cellwise regularization (as in Algorithm \ref{cellreg}) and regression (as in Algorithm \ref{crls}). 
In the first stage, given an estimate $\hat{\bm \beta}$, the loss function \eqref{cellloss3} becomes a minimisation problem with respect to $\bm \Delta$ and $\bm{\zeta}$:
\begin{equation}
 \mathop{\operatorname{argmin}}\limits_{\bm \Delta, \bm{\zeta}} \frac{1}{2}\| \bm y- ( \bm { X} - \bm \Delta) \bm {\hat\beta} - \bm {\zeta} \|_2^2 + \frac{1}{2}\|\bm X- \bm \Delta\|_F^2 + \eta\|\bm \Delta\|_1 + \theta\|\bm \zeta\|_1.
 \label{cellshrink}
\end{equation}
Note that \eqref{cellshrink} has a $L_2$ loss with $L_1$ regularization, which can be solved using the gradient descent algorithm directly. 
The cellwise regularization algorithm proceeds as in Algorithm \ref{cellreg}.

\begin{algorithm}
 \caption{Cellwise Regularization ($\hat{\bm\Delta}, \hat{\bm\zeta}\leftarrow \operatorname{CR}(\bm{ X}, \bm{y},{\hat{\bm \beta}},{\eta}, \theta)$) }
 \label{cellreg}
 \begin{algorithmic}[1] 
 \State {\bf Input: } $\bm{ X}, \bm{y},{\hat{\bm \beta}},{\eta}, \theta$
 \State {\bf Initialization: } $h \leftarrow 0$, $\bm{ \hat\Delta} ^{(0)} $ 
 \Comment {$\bm{ \hat\Delta} ^{(0)}$ is an initial estimate of $\bm{ \Delta}$}
 \Repeat 
 \State $h \leftarrow h+1$
 \State $\bm{\tilde X}^{(h)} = \bm{X} - \hat{\bm\Delta}^{(h)}$
 \State $\bm{\tilde y}^{(h)} = \bm{y} - \hat{\bm\zeta}^{(h)}$
 \State $\bm{\tilde\varepsilon}^{(h)} = \bm{\tilde y}^{(h)} - \bm{\tilde X}^{(h)}\hat{\bm\beta}$
 \State $\bm{\nabla}^{(h)} = \hat{\bm\varepsilon}^{(h)}\hat{\bm\beta}^{\top} - \bm{\tilde X}^{(h)}$
 \State $\bm{ \hat\Delta} ^{(h)} \leftarrow S\left(\bm{ \hat\Delta} ^{(h-1)} - t\bm{\nabla}^{(h)}, t \eta \right)$ 
 
 \Comment{$t$ is a stepsize, $\operatorname{S}(x, \eta) = \operatorname{sign}(x)\max \left(0, |x| - \eta\right)$}
 \State $\hat{\bm\varepsilon}^{(h)} = \bm{y} - \bm{\tilde X}\hat{\bm\beta}$
 \State $\bm{ \hat\zeta} ^{(h)} \leftarrow S\left(\hat{\bm\varepsilon}^{(h)}, \theta \right)$

 \Until $\|\hat{\bm\Delta} ^{(h)} - \hat{\bm\Delta}^{(h-1)}\|_\infty < \epsilon_1$
 \Comment{{$\epsilon_1 = 10^{-6}$} is a convergence tolerance}

 \State$\bm {\hat \Delta} \leftarrow \hat{\bm\Delta} ^{(h)}$, $\bm {\hat \zeta} \leftarrow \bm{ \hat\zeta} ^{(h)}$
\end{algorithmic}
\end{algorithm}

In the second stage, after getting $\bm {\hat \Delta}$ and $\bm {\hat \zeta}$, a regularized design matrix $\bm {\tilde X} = \bm {X} -\bm {\hat \Delta}$ is obtained as well as a regularized response $\bm{\tilde y} = \bm{y} - \hat{\bm \zeta}$. 
Then we obtain a new $\hat{\bm \beta}$ via solving
\begin{equation}
 \mathop{\operatorname{argmin}}\limits_{\bm\beta} \frac{1}{2} \| \bm {\tilde y}- \bm {\tilde X} \bm \beta \|_2^2,
 \label{ols}
\end{equation}
which is a regression loss that can be solved directly using ordinary least squares (OLS). Combining the cellwise regularization technique with OLS, we have the Cellwise Regularized Least Squares (CR-LS) to run regression and detect outliers simultaneously. 

This alternating procedure has a similarity with the EM algorithm. Initially, the algorithm detects and imputes outliers based on the observed data and current parameter estimates in the first step. In the second step, the algorithm updates the model parameters to maximize the likelihood of the observed data, incorporating the information gained from the first step.
Details of the CR-LS algorithm are as per Algorithm \ref{crls}.

\begin{algorithm}
 \caption{Cellwise Regularized Least Squares
 ($\hat{\bm\beta} 
\leftarrow \operatorname{CR-LS}(\bm{ X}, \bm{y}, {\eta}, {\theta})$) }
\label{crls} 
 \begin{algorithmic}[1] 
 \State {\bf Input: } $\bm{ X}, \bm{y},{\eta}, {\theta}$
 \State {\bf Initialization:} $\bm {\hat \beta}^{(0)}$ and $\bm {\hat \Delta}^{(0)}$, $k \leftarrow 0$ 
 \Comment {$\bm{ \hat\Delta} ^{(0)}$ is an initial estimate of $\bm{ \Delta}$}
 \Repeat 
 \State $k \leftarrow k+1$
 \State $\bm{ \hat\Delta} ^{(k)}, \hat{\bm\zeta}^{(k)} \leftarrow \operatorname{CR}(\bm{ X}, \bm{y},{\hat{\bm \beta}}^{(k-1)},{\eta}, {\theta})$ 
 \State $\bm{\tilde X}^{(k)} \leftarrow \bm{X} - \hat{\bm \Delta}^{(k)}$
 \State $\bm{\tilde y}^{(k)} \leftarrow \bm{y} - \hat{\bm \zeta}^{(k)}$
 \State $\bm{ \hat\beta} ^{(k)} \leftarrow \operatorname{OLS}(\bm{\tilde y}^{(k)}, \bm{\tilde X}^{(k)})$ 

 \Until $\|\hat{\bm\beta} ^{(k)} - \hat{\bm\beta} ^{(k-1)}\|_\infty < \epsilon_2$
 \Comment{{$\epsilon_2 = 10^{-3}$} is a convergence tolerance}
 \State $\bm{ \hat\beta} \leftarrow \bm{ \hat\beta} ^{(k)}$ 
 \end{algorithmic}
\end{algorithm}
}

\subsection{Cellwise regularized sparse regression}
For high-dimensional data and under sparse settings, only a small subset of predictors are active, which means $\bm\beta$ is a sparse vector including only a few nonzero coefficients, typically much less than $\min{(n,p)}$. Many techniques were proposed in the last thirty years to recover the sparse $\bm\beta$. For instance, the popular Lasso \citep{Tibshirani_1996} solves an $L_1$ regularized objective loss: 
\begin{equation}
 \mathop{\operatorname{argmin}}\limits_{\bm{\beta}} \frac{1}{2} \|{\bm y} - {\bm X \bm \beta}\|_2^2 + \lambda \|\bm \beta\|_1,
 \label{sparsereg}
\end{equation}
where $\lambda$ is a tuning parameter. When $\bm X$ is well-conditioned, Lasso-type estimators can guarantee a high recovery rate for $\bm \beta$ when the chosen $\lambda$ is appropriate, meaning that the estimate of $\bm \beta$ depends on $\lambda$.

To simultaneously select active variables and detect outlying cells under cellwise contamination, we combine equations \eqref{cellloss3} and 
\eqref{sparsereg} as follows:
\begin{equation}
 \mathop{\operatorname{argmin}}\limits_{\bm\beta, \bm \Delta, \bm \zeta} \frac{1}{2} \| \bm y- ( \bm { X} - \bm \Delta) \bm \beta - \bm \zeta \|_2^2 + \frac{1}{2} \|\bm X- \bm \Delta\|_F^2 + \lambda \|\bm\beta \|_1 + \eta\|\bm \Delta\|_1+ \theta\|\bm \zeta\|_1.
 \label{cellloss4}
\end{equation}

Similar to the procedure utilized to solve the optimization problem \eqref{cellloss3}, we can employ an iterative block coordinate descent algorithm to solve \eqref{cellloss4} and sequentially estimate $\bm{\beta}$, $\bm{\Delta}$, and $\bm{\zeta}$. The algorithm consists of two stages: cellwise regularization and sparse regression. In the cellwise regularization stage, given an estimate $\hat{\bm{\beta}}$, the transformed loss function remains as \eqref{cellshrink}, which can be efficiently solved using Algorithm \ref{cellreg}.

In the second stage, after getting $\bm {\hat \Delta}$ and $\bm {\hat \zeta}$, a regularized design matrix $\bm {\tilde X} = \bm {X} -\bm {\hat \Delta}$ is obtained as well as a regularized response $\bm{\tilde y} = \bm{y} - \hat{\bm \zeta}$. 
Then we obtain a new $\hat{\bm \beta}$ via solving
\begin{equation}
 \mathop{\operatorname{argmin}}\limits_{\bm\beta} \frac{1}{2} \| \bm {\tilde y}- \bm {\tilde X} \bm \beta \|_2^2 + \lambda \|\bm\beta \|_1,
 \label{lasso}
\end{equation}
which is a classical Lasso-type optimization. Combining the cellwise regularization with the Lasso, we have the Cellwise Regularized Lasso (CR-Lasso) to select variables and detect outliers simultaneously. 
Details of the CR-Lasso algorithm are stated in Algorithm \ref{crlasso}.

\begin{algorithm}[!hb]
\caption{Cellwise Regularized Lasso ($\bm{\hat\beta} 
\leftarrow \operatorname{CR-Lasso}(\bm{ X}, \bm{y},{\lambda},{\eta}, {\theta})$) } 
\hspace*{0.02in} 
{\bf Input: } $\bm{ X}, \bm{y},{\lambda},{\eta}, {\theta}$

\hspace*{0.02in} 
{\bf Initialization:} $\bm {\hat \beta}^{(0)}$ and $\bm {\hat \Delta}^{(0)}$, $k \leftarrow 0$ 
\begin{algorithmic}[1]
\Repeat
\State $k \leftarrow k+1$
\State $\bm{ \hat\Delta} ^{(k)}, \bm{\hat\zeta}^{(k)} \leftarrow \operatorname{CR}(\bm{ X}, \bm{y},{\bm{\hat \beta}}^{(k-1)},{\eta}, {\theta})$ 
\State $\bm{\tilde X}^{(k)} \leftarrow \bm{X} - \bm{\hat \Delta}^{(k)}$
\State $\bm{\tilde y}^{(k)} \leftarrow \bm{y} - \bm{\hat \zeta}^{(k)}$

\State $\bm{ \hat\beta} ^{(k)} \leftarrow \mathop{\operatorname{argmin}}\limits_{\bm\beta} \| \bm {\tilde y}^{(k)}- \bm {\tilde X}^{(k)} \bm \beta \|_2^2 + \lambda\|\bm\beta\|_1$ 

\Until $\|\bm{\hat\beta} ^{(k)} - \bm{\hat\beta} ^{(k-1)}\|_\infty < \epsilon_2$
\Comment {$\epsilon_2 = 10^{-3}$} is a convergence tolerance
\State $\bm{ \hat\beta} \leftarrow \bm{ \hat\beta} ^{(k)}$ 
\end{algorithmic}
\hspace*{0.02in} 
{\bf Output:} 
$\bm {\hat \beta}$
\label{crlasso}
\end{algorithm}

It is important to understand the convergence properties of Algorithm \ref{crlasso}. Let 
\begin{align*}
f(\bm{\beta}^{(k)}, \bm{\Delta}^{(k)}, \bm{\zeta}^{(k)}) &= \frac{1}{2} \| \bm y- ( \bm {X} - \bm{\Delta}^{(k)}) \bm{\beta}^{(k)} - \bm{ \zeta}^{(k)} \|_2^2 \\
&\quad + \frac{1}{2} \|\bm X- \bm{ \Delta}^{(k)}\|_F^2 + \lambda \|\bm{\beta}^{(k)}\|_1 + \eta \|\bm{ \Delta}^{(k)}\|_1+ \theta \|\bm{ \zeta}^{(k)}\|_1.
\end{align*}
Because the block coordinate descent iterations are non-increasing, the CR-Lasso iteration sequence $(\bm{\beta}^{(k)}, \bm{\Delta}^{(k)}, \bm{\zeta}^{(k)})$ satisfies
\begin{align*}
f(\bm{\beta}^{(k)}, \bm{\Delta}^{(k)}, \bm{\zeta}^{(k)})
\geq & f(\bm{\beta}^{(k)}, \bm{\Delta}^{(k+1)}, \bm{\zeta}^{(k)}) \\
\geq & f(\bm{\beta}^{(k)}, \bm{\Delta}^{(k+1)}, \bm{\zeta}^{(k+1)}) \\
\geq & f(\bm{\beta}^{(k+1)}, \bm{\Delta}^{(k+1)}, \bm{\zeta}^{(k+1)}),
\end{align*} 
for any $k \geq 0$.
Therefore, the final estimates converge to local minima.  {The algorithm's efficiency depends on various factors, including the dimension and the data-generating distribution. With the default convergence level set at $\epsilon_1 = 10^{-6}$  and $\epsilon_2 = 10^{-3}$} and the simulation settings detailed in {Section} \ref{simu1}, we observed convergence in all simulation runs within 20 iterations. 

\subsection{Data initialization}

{For a given dataset $[\bm{y}, \bm{X}]$, it is possible to execute \textbf{Algorithm \ref{crlasso}} with non-standardized data. However, standardizing the dataset and subsequently back-transforming the coefficients is a strategic approach to enforce equivariance properties, which would otherwise be compromised due to the regularization process.}
{For the estimation of location parameters, utilizing the median may pose a challenge, as it could fall outside the geometric boundaries of the dataset. While this concern can be mitigated by opting for the multivariate $L_1$ median \citep{dodge1999multivariate}, it comes at the expense of increased computational complexity. In this paper, we choose the classic median to enhance computational efficiency.

To achieve robust scale estimation comparable to standard deviation estimates under the normal distribution, various robust scale estimators, such as the median absolute deviation and the $Q_n$ estimator \citep{rousseeuw_alternatives_1993}, as well as the $P_n$ estimator \citep{tarr_robust_2012}, can be employed. We advocate for the use of the $Q_n$ estimator due to its robustness and efficiency.} 
This results in a robustly standardized design matrix $\bm{X}^\star$. {The effectiveness of the proposed method hinges on a suitable scale estimate, and it is imperative to determine this estimate before applying our algorithm.} We obtain a robust estimate of the residual standard deviation, $\hat\sigma$, using RLars \citep{khan2007robust}. 
Then we obtain standardized estimates $\bm{\hat\beta}^\star, \bm {\hat\Delta}^\star$, and $\bm {\hat\zeta}^\star$ from a standardized version of the objective loss:
\begin{equation}
 \mathop{\operatorname{argmin}}\limits_{\bm\beta^\star, \bm \Delta^\star, \bm \zeta^\star} \frac{1}{2} \left \| \frac{\bm y - ( \bm {X}^\star - \bm \Delta^\star) \bm \beta^\star}{\hat \sigma} - \bm \zeta^\star \right \|_2^2 + \frac{1}{2} \|\bm{X}^\star- \bm \Delta^\star\|_F^2 + \lambda \|\bm\beta^\star\|_1 + \eta \|\bm \Delta^\star\|_1+ \theta \|\bm \zeta^\star\|_1,
 \label{cellloss5}
\end{equation}
where $\bm\beta^\star, \bm \Delta^\star$, and $\bm \zeta^\star$ are the standardized version of $\bm\beta, \bm \Delta$, and $\bm \zeta$, respectively.

{Obtaining good initial values is important as the loss function \eqref{cellloss5} is non-convex.} To accelerate the algorithm's convergence and improve its accuracy, we recommend using RLars to obtain an initial estimate of $\bm{\beta}$, where the R function \texttt{rlars} is in the R package \texttt{robustHD} \citep{alfons_2021}.
{Nevertheless, in our simulation studies, we observe that even when utilizing a zero vector as the initial value for $\hat{\bm{\beta}}$, we were able to achieve satisfactory estimation results.}

\subsection{The choice of tuning parameters}
\label{subsection:tuning}

{In Lasso-type problems, a common approach for determining the optimal regularization parameters involves computing the solution path for a sequence of values of $\lambda$ and then selecting the $\lambda$ value that provides the best prediction results or smallest information criterion value. As stated by \citet{Friedman_Hastie_Tibshirani_2010}, the sequence begins with $\lambda_\text{max}$, the smallest value of $\lambda$ for which the entire vector $\bm{\hat{\beta}} = \bm{0}$. The minimum value is set as $\lambda_\text{min} = \iota \lambda_\text{max}$. Finally, a sequence of $K$ values for $\lambda$ is constructed, increasing logarithmically from $\lambda_\text{min}$ to $\lambda_\text{max}$. In our studies we set $\iota = 0.001$ and $K = 50$.}

Two of the classical selection criteria for sparse regression models are the AIC \citep{akaike1974new} and BIC \citep{Schwarz_1978}, where $\mathrm{AIC} = L+2k$, $\mathrm{BIC} = L+\log(n)k$ and $L$ denotes the corresponding loss and $k$ is the number of active predictors. 
For the proposed method, we define
\begin{equation}
 L = 
 \left \| \frac{\bm{ y}- ( \bm {X}^\star - \bm {\hat\Delta^}\star) \bm {\hat\beta}^\star}{\hat \sigma} - \bm {\hat\zeta}^\star \right \|_2^2 + 2\theta \|\bm {\hat\zeta}^\star\|_1.
\end{equation}
We use the BIC as the default selection criterion because of its ease of implementation and good performance in our simulation studies and real data applications. For conciseness, we do not report results for criteria other than the BIC. 

We implement a hard threshold for model selection to avoid potential over-regularisation issues, where the algorithm selects only a few predictors and shrinks almost all the cells in the selected predictors. {Such an occurrence is more likely when setting a relatively large $\lambda$. To address this, we exclude any model from consideration when the tuning parameter $\lambda$ results in a cell shrinking rate in one active predictor that exceeds $30\%$ of the number of observations.} 

Regarding $\eta$, a natural choice is to set $\eta = z_{0.995}$ for all cells, where $z_{0.995} = 2.576$, the $99.5\%$ quantile of the standard normal distribution. A quantile threshold is commonly used, such as in Huberization \citep{huber_robust_2004} and DDC \citep{rousseeuw_detecting_2018}. 
Similarly, setting $\theta = z_{0.995}$ is also a natural choice. However, this parameter is sensitive to the estimated error scale, $\hat\sigma$, which may not be sufficiently accurate under cellwise contamination. Therefore, to ensure robustness in our simulation studies, we set $\theta = 1$. 
\citet{maronna2019robust} showed that good efficiency under Gaussian assumptions can be achieved despite a conservative choice for $\theta$.

{
\subsection{Post-cellwise-regularized regression}
It is well known that the Lasso estimator can introduce bias. To counteract this, we perform CR-Lasso with post-cellwise-regularized regression based on the selected predictors to enhance the model's performance. The process is as follows:

\begin{enumerate}
    \item Parameters $\eta$ and $\theta$ are established as mentioned in Section \ref{subsection:tuning}. This choice is made since optimising three tuning parameters simultaneously is time-consuming in practice.
    \item For a grid of $\lambda$ values, we run Algorithm \ref{crlasso} sequentially and use BIC to choose the optimal tuning parameter $\lambda_\text{opt}$.
    \item After obtaining $\lambda_\text{opt}$, we run Algorithm \ref{crls} using only the variables that were retained under $\lambda_\text{opt}$ to obtain the final regression coefficient vector estimate $\bm{\hat{\beta}}$.
\end{enumerate}

Our simulation studies presented in Section \ref{simu} consider the performance of CR-Lasso with post-cellwise-regularized regression. In \ref{simu_post}, we show the performance of CR-Lasso both with and without post-cellwise-regularized regression. The findings indicate that even without post-cellwise-regularized regression, CR-Lasso still outperforms other methods under cellwise contamination.}

\section{Simulation}
\label{simu}
To demonstrate the effectiveness of the proposed method, we ran simulation studies and compared the performance of six methods for a moderate-dimensional and a high-dimensional setting: CR-Lasso, sparse shooting S \citep[SSS]{Bottmer_Croux_Wilms_2022}, robust Lars \citep[RLars]{khan2007robust}, MM-Lasso \citep{Smucler_Yohai_2017}, sparse LTS \citep[SLTS]{alfons_sparse_2013} and Lasso \citep{Tibshirani_1996}.
For the implementation of SSS, we utilized the R function \texttt{sparseshooting} \citep{Wilms_2020}. RLars and SLTS were implemented using the R functions \texttt{rlars} and {\texttt{sparseLTS}} in the R package \texttt{robustHD}  \citep{alfons_2021}. The R function \texttt{mmlasso} \citep{Smucler_2017} was employed for MM-Lasso. Lasso was implemented through the R function \texttt{glmnet} in the R package \texttt{glmnet}  \citep{Friedman_Hastie_Tibshirani_2010}.
{By default in the utilized functions, Lasso and MM-Lasso employ cross-validation (CV) to select optimal tuning parameters. In contrast, SSS, RLars, and CR-Lasso utilize the Bayesian information criterion (BIC) for tuning parameter selection. SLTS uses a default tuning parameter without optimization.}
Among all the compared methods, CR-Lasso, SSS, and RLars are relatively robust to cellwise outliers, MM-Lasso and SLTS are rowwise robust methods, and Lasso is a classical variable selection technique that is not robust to any outliers and is included here for completeness. 



\subsection{Moderate-dimensional setting}
\label{simu1}
In our simulations, we set $n = 200$, $p = 50$ and $\bm\beta = (\bm 1_{10} ^\top, \bm 0_{p-10}^\top )^\top$ {with an intercept term $\beta_0 = 1$}. Clean observation vectors $\bm{\check x}_\mathrm{i}$ were sampled from $N(\bm 0,\bm\Sigma)$, and errors $\varepsilon_i$ were sampled from $N(0,3^2)$, indicating a relatively high level of noise in the data. The correlation structure among predictors was given by $\Sigma _{ij} = \rho^{|i-j|}$ and we set $\rho = 0.5$. We also generated $\bm{\check x}_i$ from the multivariate $t_4$ distribution to simulate distributions with heavier tails.

To introduce cellwise outliers, we set the contamination proportion $e$ to $ 0\%$, $ 2\%$ and $5\%$ for all predictors and generate outliers independently. Outlying cells $\Delta_{ij}$ and {$\zeta_{i}$} were randomly generated from $N(\gamma, 1)$ and $N(-\gamma, 1)$ with equal probability, where $\gamma$ varies to simulate outliers with different magnitudes. 

We ran $200$ simulations for each scenario and used the root of the mean squared prediction error (RMSPE) to assess the prediction accuracy of the considered methods.
In addition, to assess the accuracy of variable selection, we employed
\begin{equation}
 \mathrm{F}_1 = \frac{2\mathrm{TP}}{2\mathrm{TP} + \mathrm{FP} + \mathrm{FN}},
\end{equation}
where TP, FP, and FN indicate true positives, false positives, and false negatives, respectively. 
While the $\mathrm{F}_1$ score is commonly used for classification problems, it is also used to evaluate the performance of variable selection techniques, as in \citet{bleichvariable2014}.

The advantage of using the $\mathrm{F}_1$ score to measure variable selection is that it takes into account both the precision and recall of the selected variables. Precision measures the proportion of selected variables that are relevant, while recall measures the proportion of relevant variables that are selected. By combining precision and recall, the $\mathrm{F}_1$ score provides a balanced evaluation of variable selection performance.
This allows us to measure the effectiveness of each method in selection and prediction.

\begin{figure}[!ht]
	\centering
	\includegraphics[width=12cm]{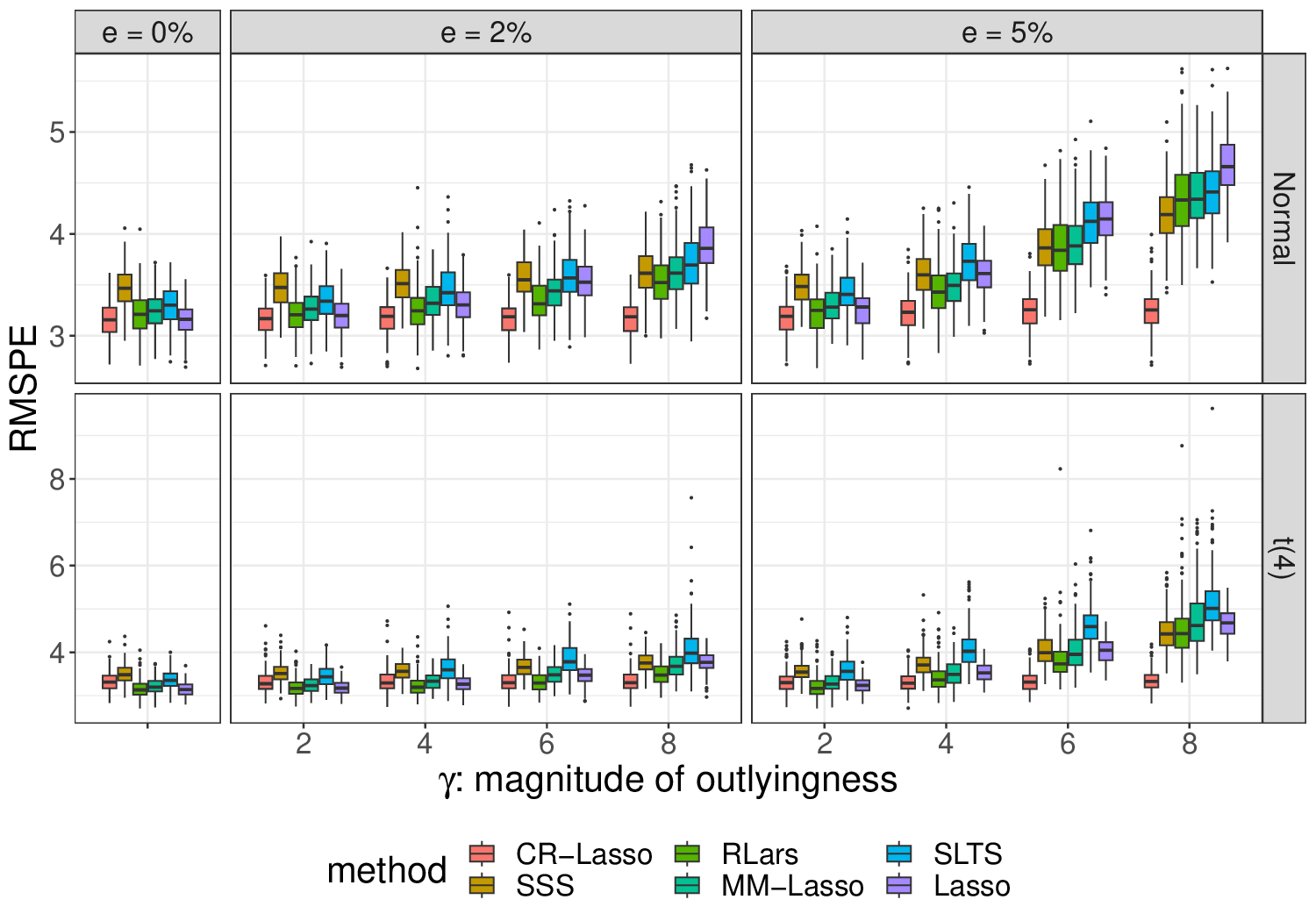}
	\caption{Prediction results as summarised by the $\text{RMSPE}$ over 200 simulation runs for various contamination rates when $p = 50$, from which $10$ predictors are active. Each column represents a contamination rate $e$, and each row represents a distribution type of predictor.}
	\label{rmspe50}
\end{figure}

Figure \ref{rmspe50} reports the distributions of RMSPEs for the data generated with $p = 50$. 
When the predictors are generated from a normal distribution, as depicted in the top row of the figure, CR-Lasso demonstrates superiority compared to other methods.
Without contamination ($e = 0\%$), CR-Lasso, RLars, MM-Lasso and Lasso exhibit similar performance, with an average RMSPE of around $3.2$, while SLTS and SSS display inferior performance.
At a $2\%$ contamination rate, CR-Lasso exhibits superior and stable performance, even with a high magnitude of outlyingness. On the other hand, Lasso, MM-Lasso, RLars, SLTS and SSS perform inferior with increasing $\gamma$. 
At a $5\%$ contamination rate, CR-Lasso maintains stable performance. The prediction results are similar to $2\%$ contamination. In contrast, other methods show inferior results compared with their counterparts under $2\%$ contamination.


The behaviour of the six compared methods differs significantly when the predictors follow a $t_4$ distribution, which is known to generate many high-leverage points. CR-Lasso exhibits inferior performance compared to the normal cases since it treats high-leverage points as outliers and shrinks their values, resulting in biased estimates. However, even in this challenging scenario, CR-Lasso outperforms other methods with $2\%$ or $5\%$ contamination with high magnitudes of outlyingness when $\gamma$ is as high as $6$ or $8$. The overall performance of all compared methods is similar to the Gaussian setting, except for some extreme RMSPE values observed in MM-Lasso, RLars, and SLTS.

\begin{figure}[!ht]
	\centering
	\includegraphics[width=12cm]{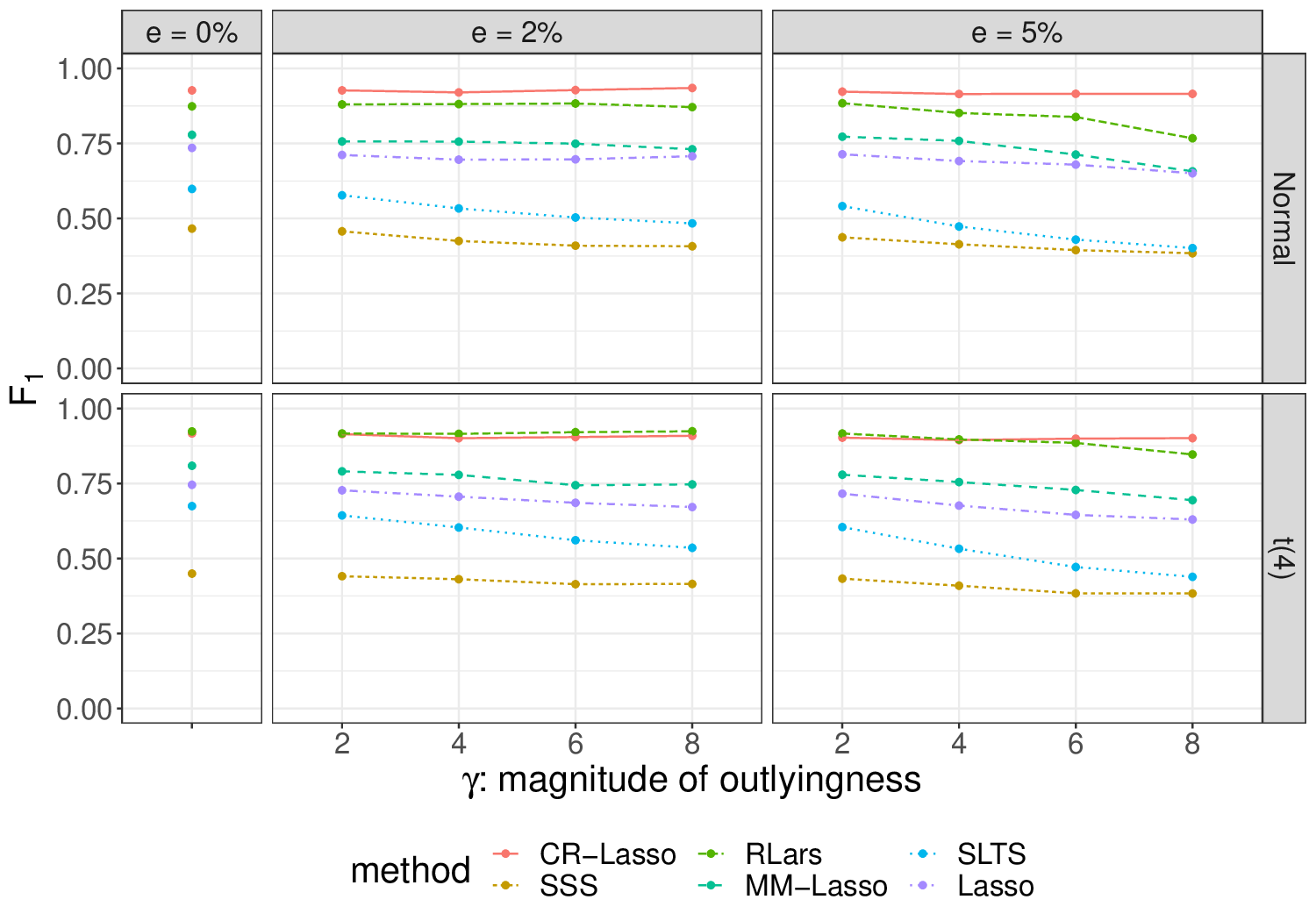}
	\caption{Variable selection results as summarised by the $\mathrm{F}_1$ scores over 200 simulation runs for various contamination rates when $p = 50$, from which $10$ predictors are active. Each column represents a contamination rate $e$, and each row represents a distribution type of predictor.}
	\label{f1_50}
\end{figure}

Figure \ref{f1_50} presents the mean $\mathrm{F}_1$ scores across all evaluated methods. When the predictors are generated from a normal distribution, as depicted in the top row of the figure, CR-Lasso demonstrates exceptional $\mathrm{F}_1$ scores across all scenarios. RLars also exhibits good $\mathrm{F}_1$ scores, followed by MM-Lasso and Lasso. On the other hand, SLTS and SSS exhibit inferior $\mathrm{F}_1$ scores as they select many inactive variables. 
When the predictors follow a $t_4$ distribution, CR-Lasso does not perform as well as when the predictors are multivariate normal. RLars shows competitive results as CR-Lasso since there are more good high leverage points, strengthening the estimate. The overall performance of other compared methods for heavy-tailed predictors is similar to the normal setting.

\subsection{High-dimensional setting}
\label{simu2}
In this subsection, we present the simulation results for high-dimensional settings. Specifically, we set the number of predictors to $p=300$ and $\bm \beta = (\bm 1_{10}^\top, \bm 0_{p-10}^\top)^\top$, keeping all other settings the same as in the moderate-dimensional settings.

\begin{figure}[!ht]
	\centering
	\includegraphics[width=12cm]{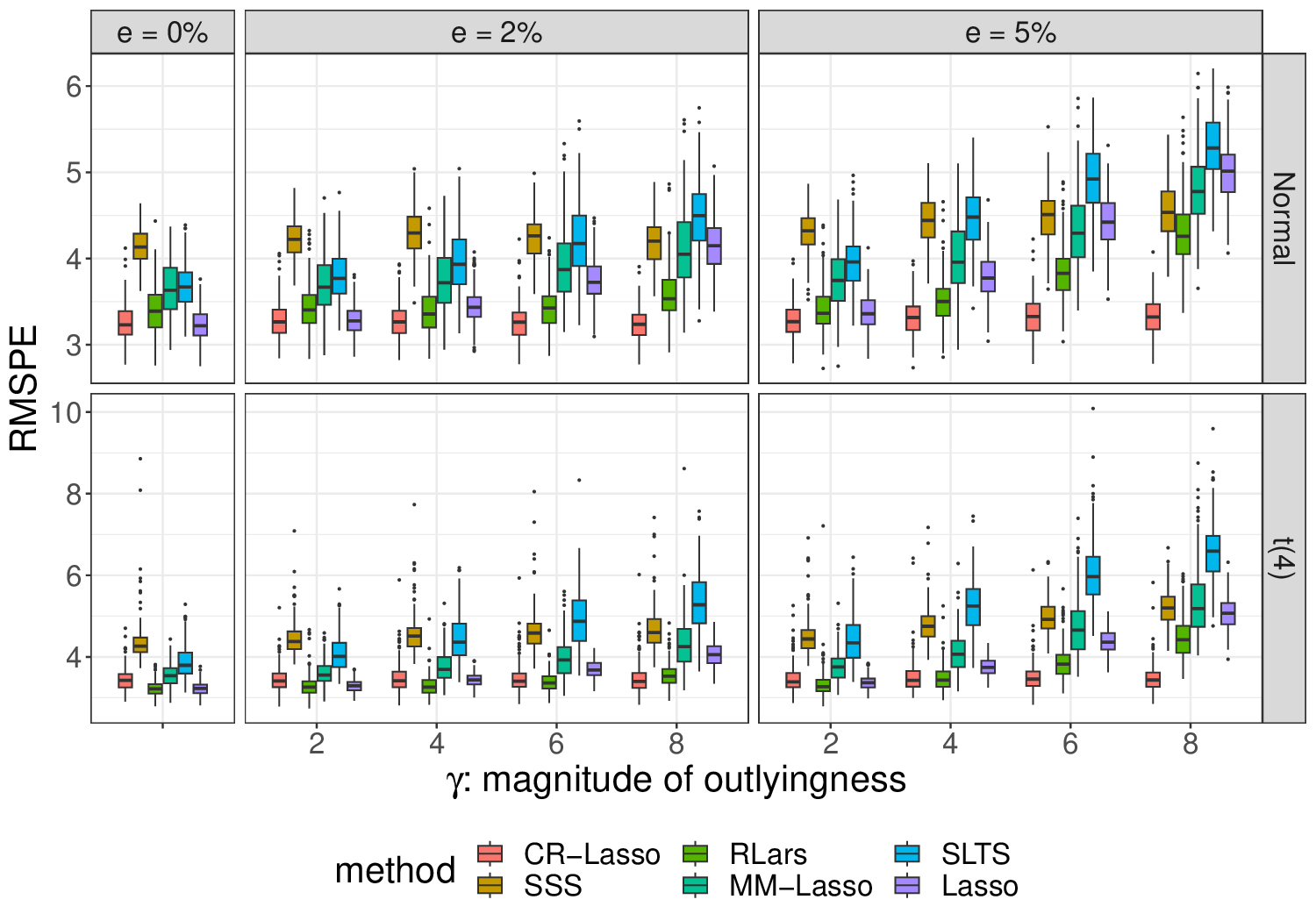}
	\caption{Prediction results as summarised by the $\text{RMSPE}$ over 200 simulation runs for various contamination rates when $p = 300$, from which $10$ predictors are active. Each column represents a contamination rate $e$, and each row represents a distribution type of predictor.}
	\label{rmspe300}
\end{figure}

Figure \ref{rmspe300} reports the results of RMSPEs for the data generated with $p = 300$. With such a high noise level, all compared methods exhibit significantly different performances in high-dimensional cases compared to their moderate-dimensional counterparts. Specifically, with $e=0\%$ or $2\%$ contamination, only CR-Lasso and RLars demonstrate considerable performance, while Lasso also shows superior performance with a low magnitude of outlyingness. MM-Lasso, SLTS and SSS perform less effectively.
When the contamination rate increases to $5\%$, only CR-Lasso maintains stable performance. Other methods show inferior performance in this case, particularly with the increase of $\gamma$.

Like the moderate-dimensional cases, CR-Lasso performs better with normally distributed predictors than when the predictors follow a multivariate $t_4$ distribution. In contrast, RLars exhibits better performance because of the presence of good high-leverage points. Even in this case, CR-Lasso still performs better with a $5\%$ contamination rate and a large magnitude of outliers. Besides, MM-Lasso, SLTS and SSS exhibit many extreme RMSPE values, indicating that they are not robust to outliers with such a high-level noise.

\begin{figure}[!ht]
	\centering
	\includegraphics[width=12cm]{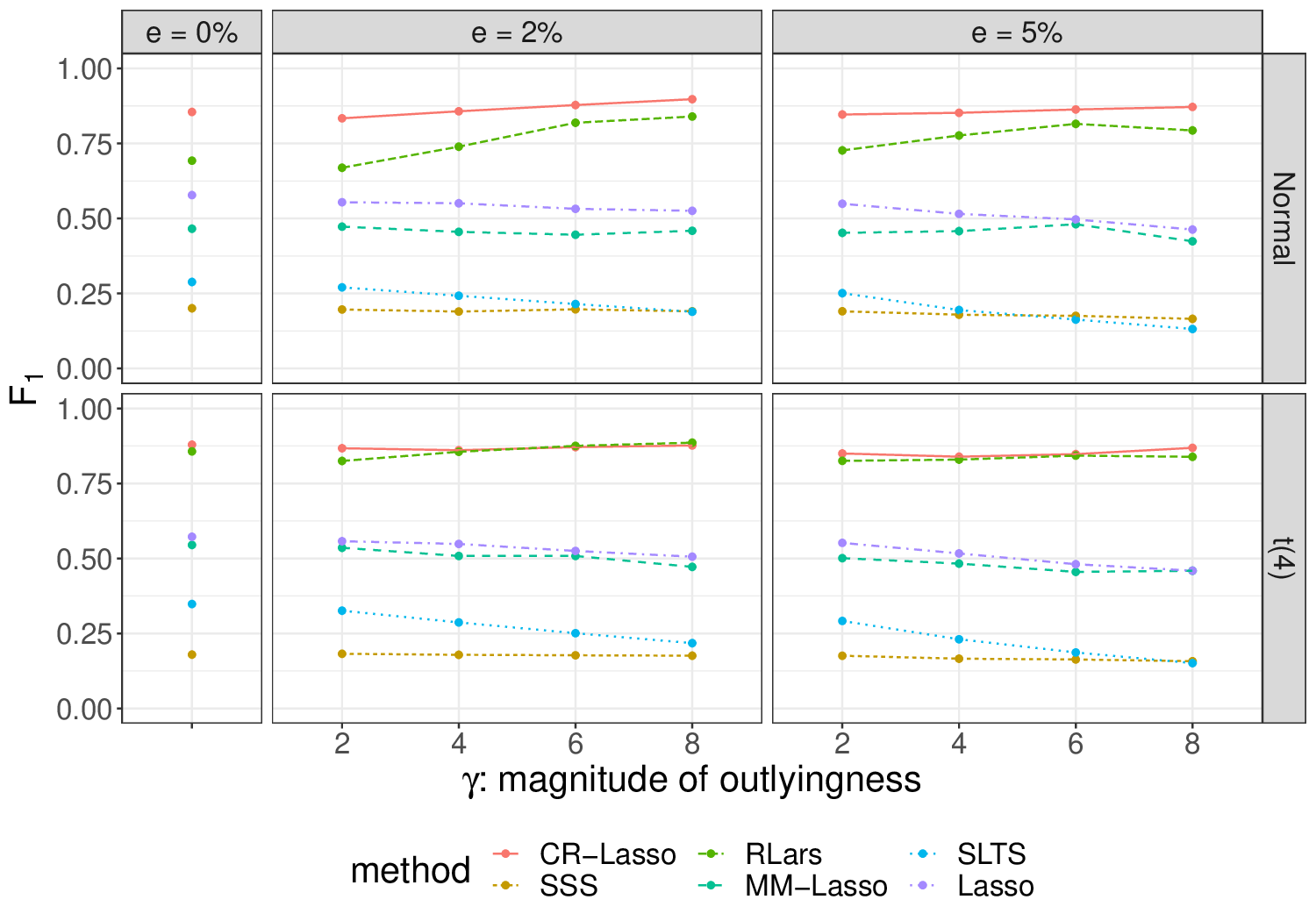}
 \caption{Selection results as summarised by the $\mathrm{F}_1$ scores over 200 simulation runs for various contamination rates when $p = 300$, from which $10$ predictors are active. Each column represents a contamination rate $e$, and each row represents a distribution type of predictor.}
 \label{f1200}
\end{figure}

Figure \ref{f1200} depicts the mean $\mathrm{F}_1$ scores across all evaluated methods in the high-dimensional cases. Compared to the moderate-dimensional cases, all the methods demonstrate inferior $\mathrm{F}_1$ scores. 
RLars exhibits competitive $\mathrm{F}_1$ scores with a $t_4$ distribution. Conversely, Lasso, MM-Lasso, SLTS and SSS show poor $\mathrm{F}_1$ scores as they tend to select too many inactive predictors.

\subsection{Additional simulation results}
{Additional simulation results are presented in the Appendix. 
First, Appendix A demonstrates commendable performance of CR-Lasso even in the absence of post-regression, although it may not reach the same level of superiority observed when post-regression is applied. Second, Appendix B demonstrates the effectiveness of CR-Lasso under rowwise contamination. Moreover, it is noteworthy, however, that CR-Lasso, like other methods, faces challenges when dealing with predictors generated from heavy tail distributions, such as from Cauchy distribution (see Appendix C).}
\section{The bone mineral density data}
\label{data}

To illustrate the proposed methodology, we considered the bone mineral density (BMD) data from \citet{reppe2010eight}. {The data are publicly available in the European Bioinformatics Institute Array-Express repository (https://www.ebi.ac.uk/arrayexpress/experiments/E-MEXP-1618/)}.
The BMD dataset consists of gene expression measurements of 54,675 probes of 84 Norwegian women.
Microarray measurements are often contaminated (noisy), as \citet{rocke2001model} highlighted. This contamination can stem from multiple sources, thereby obscuring the gene expression in the data \citep{zakharkin2005sources}.

Given the large number of variables in the dataset, a pre-screening step was implemented to identify the subset of variables that are most correlated with the outcome of interest, the total hip T-score. To accomplish this, we first log-transformed all the predictors and then utilized the robust correlation estimate based on winsorization as in \citet{khan2007robust}, instead of the Pearson correlation, since winsorization is more robust to outliers that may occur in the dataset. 
The screened data comprise measurements of $p = 100$ genes from $n = 84$ Norwegian women. 

\begin{figure}[!ht]
	\centering
	\includegraphics[width=12cm]{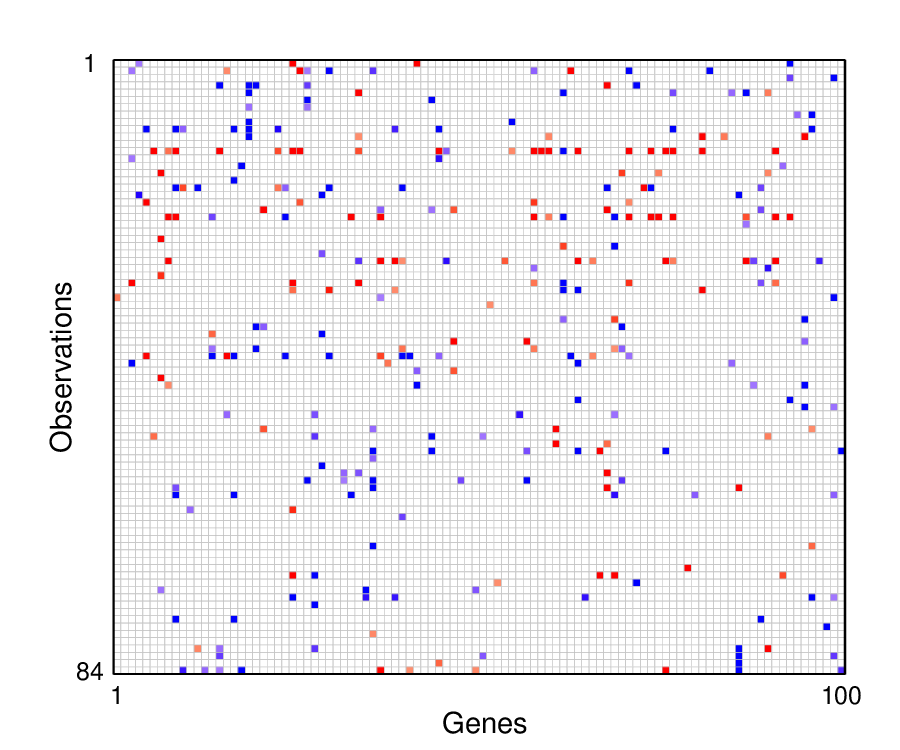}
	\caption{Outlier cell map for 100 screened variables on 84 Norwegian women. Cells are flagged as outlying if the observed and predicted values differ too much. Most cells are blank, showing they are not detected as outliers. A red cell means the observed value is significantly higher than the predicted value, and a blue cell means the observed value is significantly lower.}
	\label{tbone}
\end{figure}

Figure \ref{tbone} displays the outlier detection results based on the screened predictors using DDC \citep{rousseeuw_detecting_2018}. 
On average, the screened genes exhibit a contamination rate of $3.61\%$, with the probe $\textrm{236831\_at}$ having the highest contamination rate of $9.52\%$. 
Among the observations, the thirteenth observation{, identified by patient ID 20,} shows the highest contamination rate at $22\%$.

Given the cellwise contamination, we first standardized all variables with the median and $Q_n$. We then conducted a simple simulation study to validate the effectiveness of CR-Lasso, Lasso, MM-Lasso, RLars, SLTS and SSS on the bone mineral density data. 

We first obtained a clean (imputed) dataset $\bm{\check X}$ using DDC \citep{rousseeuw_detecting_2018}. We then  generated an artificial response $\bm{y} = \bm{\check X}\bm{\beta} + \bm{\varepsilon}$ using screened clean predictors and $\bm{\varepsilon} \sim N(\bm{0},0.5^2\bm{I})$. We randomly picked ten active predictors in each simulation run and set $\beta_j \sim U(1, 1.5)$ for each of them. We then randomly collected 80\% observations from the original (contaminated) dataset for model training, while the remaining 20\% of the imputed (clean) dataset was used to assess the prediction accuracy. We repeated the simulation procedure 200 times.  For each method, we show MAPE (mean absolute prediction error) and RMSPE values in Table \ref{tbonerstsimu}, as well as their True positive numbers (TP), true negative numbers (TN) and F$_1$ scores.

\begin{table}[!ht]
\centering
\caption{The average of RMSPEs, MAPEs, TPs, TNs and F$_1$s from 200 simulations. The method that yields the highest performance is highlighted in bold.}
\resizebox{12cm}{!}{
\begin{tabular}{lrrrrrr}
 \hline
 & CR-Lasso & SSS & RLars & MM-Lasso & SLTS & Lasso \\ 
 \hline
 RMSPE & \textbf{1.43} & 2.40 & 1.73 & 1.91 & 2.75 & 1.64 \\ 
 MAPE & \textbf{1.16} & 1.94 & 1.40 & 1.55 & 2.22 & 1.33 \\ 
 TP & 8.66 & 6.55 & 7.92 & 7.84 & 6.14 & \textbf{9.31} \\ 
 TN & 76.26 & 75.88 & \textbf{77.05} & 73.49 & 75.75 & 67.83 \\ 
 F$_1$ & \textbf{0.55} & 0.45 & 0.52 & 0.48 & 0.41 & 0.46 \\ 
 \hline
\end{tabular}}
\label{tbonerstsimu}
\end{table}

Table \ref{tbonerstsimu} shows a similar conclusion to the simulation results. CR-Lasso shows superior performance over the other methods considered, evidenced by having the lowest average RMSPE and MAPE values. The Lasso method also performs reasonably well on this dataset. This is expected as there are no extreme outliers in this data and the fraction of cellwise outliers is modest. 

To demonstrate the performance on the real data, a sparse regression model was fitted separately with the original response (the total hip T-score) via the aforementioned methods. Model selection results (model sizes) are shown in Table \ref{tbonerst}. Out of the 100 pre-screened genes, nine genes were commonly chosen by CR-Lasso, Lasso, MM-Lasso, RLars, and SLTS. We ran the Leave-One-Out Cross-Validation to assess the performance of the selected models. The RMSPE and MAPE (mean absolute prediction error) from Leave-one-out prediction residuals are also shown in Table \ref{tbonerst}.

\begin{table}[!ht]
\centering
\caption{Model size, RMSPE and MAPE from Leave-one-out prediction residuals. The method that yields the highest performance is highlighted in bold.}
\resizebox{12cm}{!}{
\begin{tabular}{lrrrrrr}
 \hline
 & CR-Lasso & SSS & RLars & MM-Lasso & SLTS & Lasso \\ 
 \hline
 Model size & 25 & 11 & 20 & 18 & 25 & 27\\
 RMSPE & 1.23 & 1.55 & 1.48 & 1.15 & 1.25 & \textbf{0.96} \\ 
 MAPE & 0.98 & 1.25 & 1.20 & 0.95 & 1.06 & \textbf{0.79} \\ 
 \hline
\end{tabular}}
\label{tbonerst}
\end{table}

From Table \ref{tbonerst}, it is clear that there are significant differences in the prediction outcomes obtained by the various methods evaluated. For this real data, Lasso outperforms other methods by exhibiting the lowest RMSPE and MAPE values, this is followed by the MM-Lasso and CR-Lasso. We note that for this data, RLars and SSS have slightly larger prediction errors, highlighting that in this example, these cellwise-robust methods would require at least some minimal amount of extreme cellwise outliers to show superior prediction performance.

\section{Discussion}
\label{discussion}

Cellwise outliers can create significant challenges when building a regression model. Most observations may eventually be contaminated by outliers in regression models due to the propagation of cellwise outliers. Most of the existing methods may not effectively identify and manage such outliers. Therefore, there is a need to identify and manage these outliers.

Motivated by this challenge, we proposed a novel approach called CR-Lasso, which incorporates a constraint on the deviation of each cell in the loss function to detect cellwise outliers in regression models. We developed an iterative procedure for sparse regression and outlier detection by combining Lasso and cellwise outlier regularisation. Our simulation studies and real data analysis demonstrate that CR-Lasso generally has superior variable selection performance and estimation accuracy compared to other methods when outliers are present, especially when the noise ratio is high and the magnitudes of outliers are extreme. In datasets with only a few outliers that are not too extreme, traditional non-robust estimators, such as the Lasso, perform considerably well.

However, we acknowledge two critical issues that require further investigation. Firstly, estimating $\sigma$ accurately under cellwise contamination is challenging. Although we used RLars \citep{khan2007robust} to obtain an initial estimate, it can overestimate $\sigma$ when the noise ratio is high. Therefore, obtaining a more precise initial estimate of $\sigma$ is necessary and needs further exploration.
Secondly, we only considered symmetric light-tailed predictors in this paper. However, handling asymmetric or heavy-tailed predictors with outliers is challenging, as the algorithm may identify clean tails as outliers. {Our simulation results indicate that CR-Lasso may have inferior performance under extremely heavy-tailed distributions such as Cauchy and $t_1$.} In real data applications, it is recommended to perform data transformations as a preprocessing step to improve the model performance. For instance, we applied the log transformation in our real data application. Other transformations such as the Box-Cox transformation \citep{box1964analysis} may also be beneficial. Nevertheless, it is crucial to conduct further investigation to develop more robust methods that can handle asymmetric or heavy-tailed predictors effectively.
{
Moreover, the loss function \eqref{cellloss3} can be formulated as a constrained mixed-integer optimization problem, providing an opportunity for resolution through alternative algorithms, similar to the work of \citet{insolia2022simultaneous}, which can be explored in the future.

In conclusion, our proposed CR-Lasso approach produces competitive results. Meanwhile, addressing the above-mentioned issues is essential for robust outlier detection and feature selection in more general settings.
}


\section*{Acknowledgements}

Su's research was supported by the Chinese Scholarship Council \newline
\#{}201906360181.
Muller and Tarr's research was supported by the Australian Research Council Discovery Project \#{}210100521.
Wang’s research was supported in part by the Sydney Mathematical Research Institute, the University of Sydney, and the Australian National University International Visitor Program.

\newpage
\appendix

\section{Simulation results without post-regression}
\label{simu_post}

\begin{figure}[!ht]
	\centering
	\includegraphics[width=12cm]{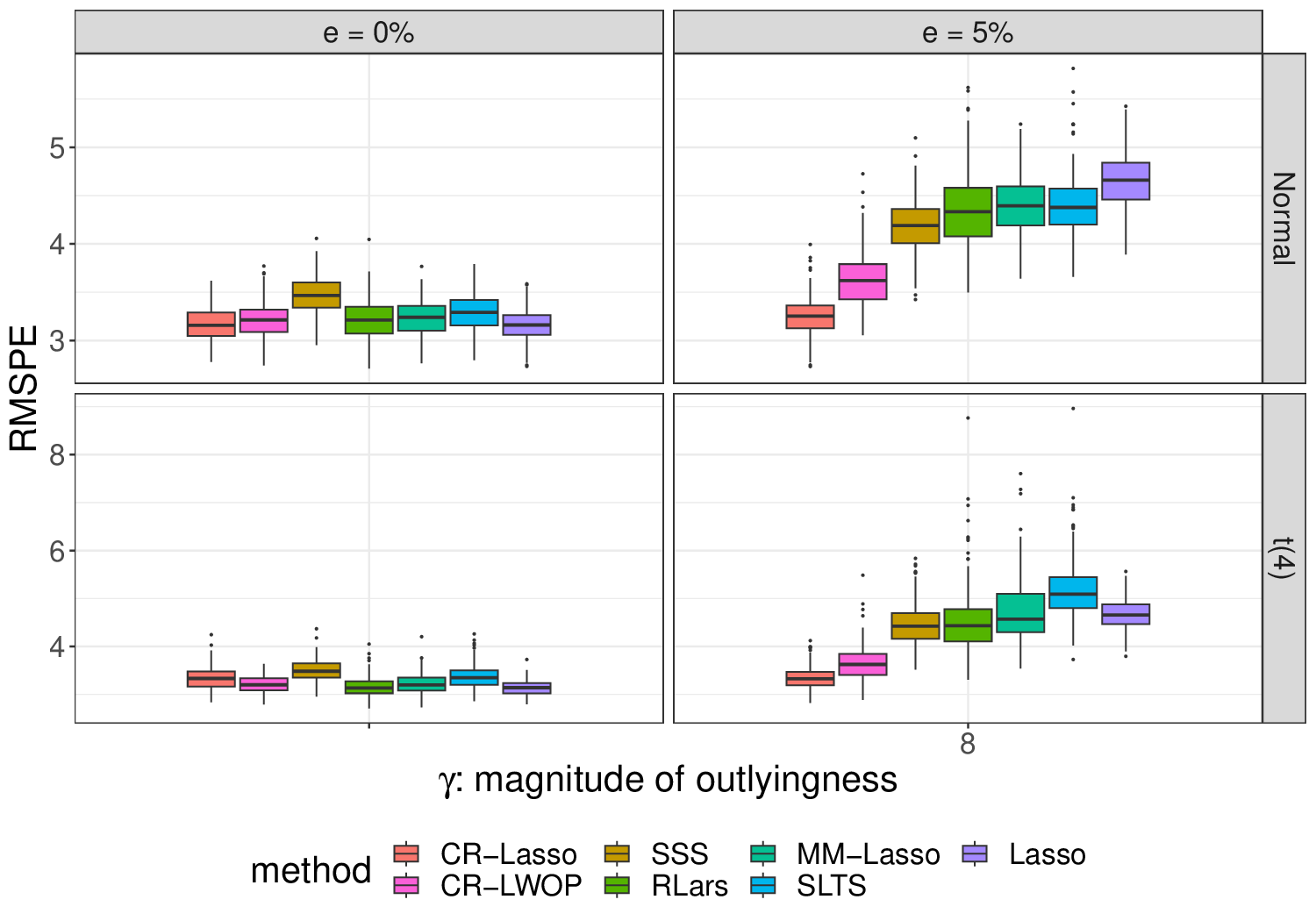}
	\caption{Prediction results as summarised by the {$\mathrm{RMSPE}$} over 200 simulation runs for various contamination rates when $p = 50$, from which $10$ predictors are active. Each column represents a contamination rate $e$, and each row represents a distribution type of predictor.}
	\label{rmspe_post}
\end{figure}

To assess the performance of CR-Lasso with and without post-cellwise-regularized regression, we conducted additional simulations. For simplicity, we adopted the same configuration as outlined in Section~\ref{simu1}, where only $e$ is set to $0\%, 5\%$, and $\gamma = 8$. In this context, we contrast the performance of CR-Lasso (Cellwise Regularized Lasso with post-cellwise-regularized regression), CR-LWOP (Cellwise Regularized Lasso WithOut Post-regression), and other methods. Figure \ref{rmspe_post} illustrates the distributions of RMSPEs derived from the simulation results.

Figure \ref{rmspe_post} suggests that post-cellwise-regularized regression contributes to enhanced prediction capabilities. Even without post-regression, CR-Lasso exhibits superior performance compared to other methods. It is crucial to note that post-regression is solely employed to enhance prediction and does not influence variable selection results. Consequently, the variable selection outcomes for CR-Lasso with or without post-regression remain unchanged.

\section{Simulation under rowwise regression}
\label{simu_rowwise}

\begin{figure}[!htp]
	\centering
	\includegraphics[width=12cm]{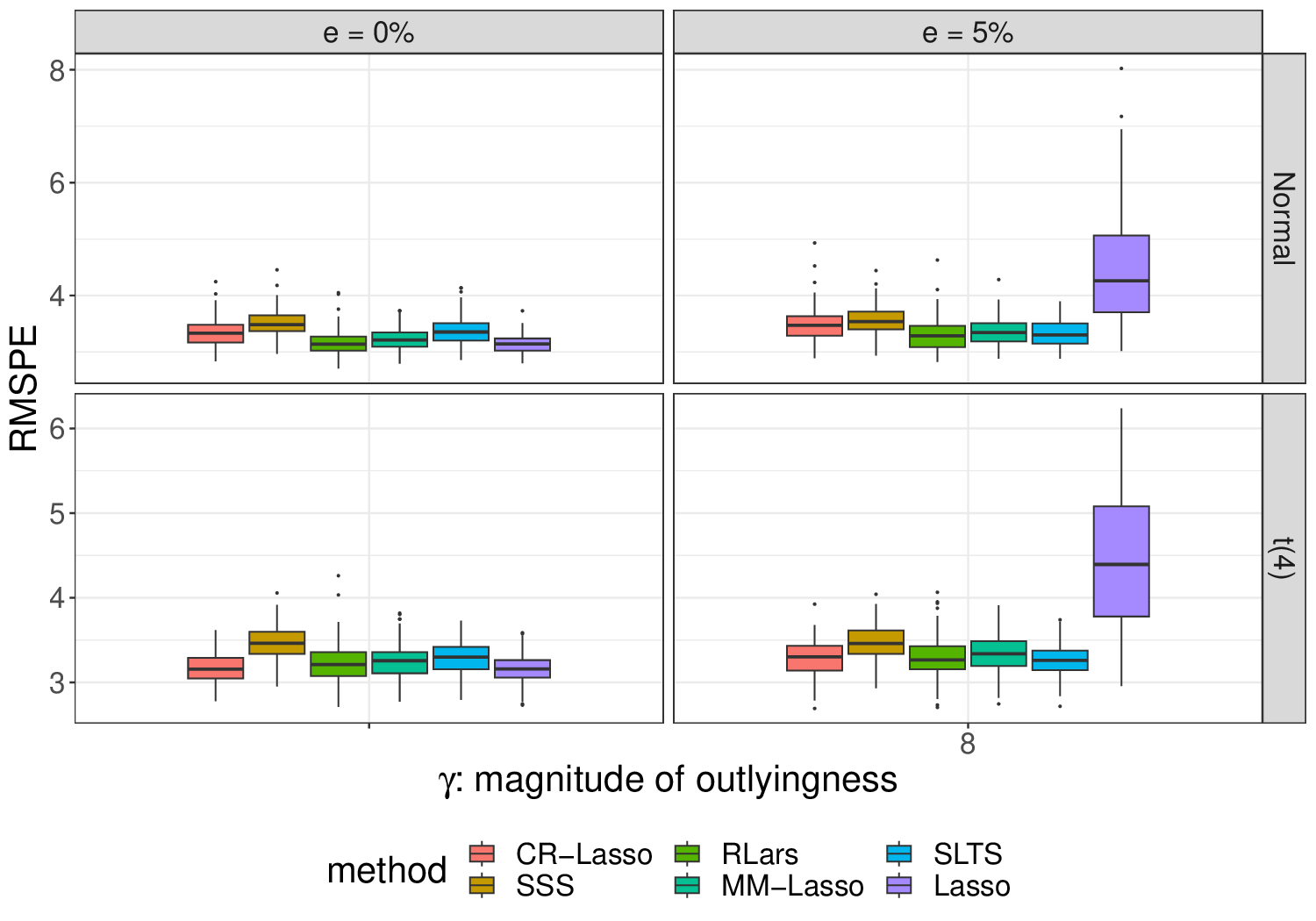}
	\caption{Prediction results as summarised by the $\text{RMSPE}$ over 200 simulation runs for various contamination rates when $p = 50$, from which $10$ predictors are active. Each column represents a contamination rate $e$, and each row represents a distribution type of predictor.}
	\label{rmspe_rowwise}
\end{figure}

To demonstrate the performance of the proposed method in broader contexts, we extended the comparison to include rowwise contamination. {The simulation settings were the same as in \ref{simu_post}, with the exception that outliers are generated rowwise. That is, outlying cells were still randomly generated from $N(8, 1)$ and $N(-8, 1)$ with equal probability, while the distinction lies in the fact that cells within the same observation were simultaneously contaminated or remained clean.} Figure \ref{rmspe_rowwise} presents the distributions of RMSPEs derived from the simulation results. Table \ref{rowwise_f1} shows the variable selection results of the compared methods. 

\begin{table}[!htp]
\centering
\caption{Mean F$_1$ from 200 simulations. The method that yields the highest performance is highlighted in bold.}
\resizebox{12cm}{!}{
\begin{tabular}{llrrrrrr}
\hline
e & df & CR-Lasso & SSS & RLars & MM-Lasso & SLTS & Lasso \\ 
\hline
 e = 0\% & Normal & 0.91 & 0.45 & $\mathbf{0.93}$ & 0.80 & 0.68 & 0.74 \\ 
         & t(4) & $\mathbf{0.92}$ & 0.47 & 0.87 & 0.77 & 0.60 & 0.73 \\ 
 e = 5\% & Normal & 0.85 & 0.43 & $\mathbf{0.89}$ & 0.72 & 0.65 & 0.65 \\ 
         & t(4) & $\mathbf{0.85}$ & 0.43 & $\mathbf{0.85}$ & 0.69 & 0.59 & 0.61 \\ 
\hline
\end{tabular}}
\label{rowwise_f1}
\end{table}

{The findings from Figure \ref{rmspe_rowwise} and Table \ref{rowwise_f1} suggest that CR-Lasso demonstrates competitive performance in the context of rowwise contamination. All methods exhibit similar {RMSPEs} without contamination regardless of the data-generating distributions, while SLTS shows slightly worse performance. When subjected to $5\%$ contamination, all methods, except Lasso, deliver considerable prediction results. This is attributed to the robustness of the compared methods in handling rowwise outliers.}

{
\section{Simulation under Cauchy distribution}
\label{simu_cauchy}
We also did simulations where predictors were generated from the standard Cauchy  (i.e., the $t_1$) distribution, while the rest simulation settings were the same as in \ref{simu_post}.}

\begin{figure}[!htp]
	\centering
    
	\includegraphics[width=12cm]{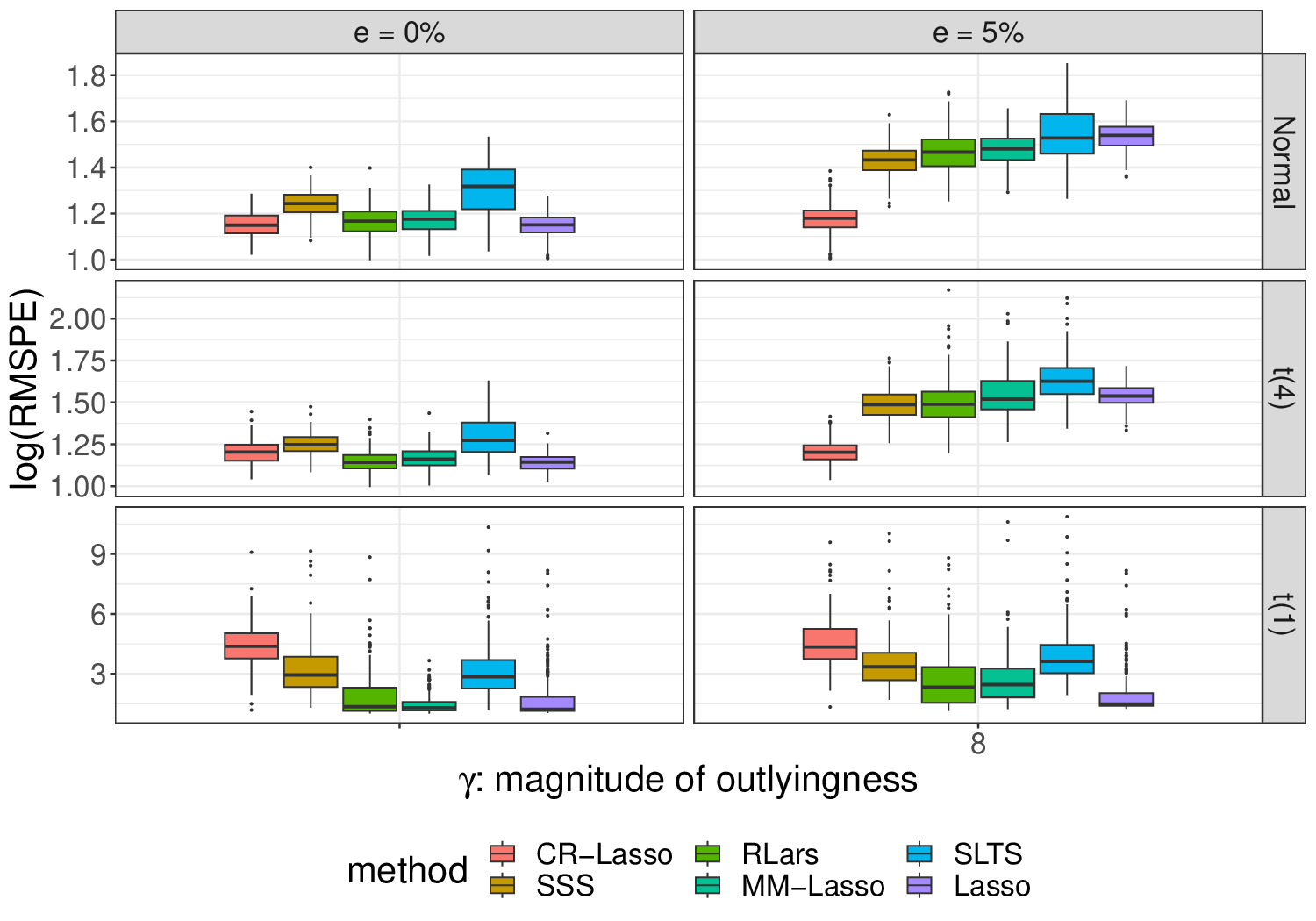}
	\caption{Prediction results as summarised by the {$\log(\text{RMSPE})$} over 200 simulation runs for various contamination rates when $p = 50$, from which $10$ predictors are active. Each column represents a contamination rate $e$, and each row represents a distribution type of predictor.}
	\label{rmspe_cauchy}
\end{figure}

Considering {RMSPEs} (log-transformed) depicted in Figure \ref{rmspe_cauchy}, it becomes evident that all methods display noticeable extreme outliers when predictors are derived from the Cauchy distribution. Meanwhile, CR-Lasso shows inferior performance as it does not have a good strategy to distinguish outliers and good leverage points. Nevertheless, it is important to note that the Cauchy distribution represents an extreme case and is likely unrealistic in practical applications. 




%
%
%
\newpage
\bibliographystyle{elsarticle-harv}
\bibliography{crlasso}

\end{document}